\documentclass{article}

     \usepackage[final, nonatbib]{neurips_2021}

\usepackage[utf8]{inputenc} 
\usepackage[T1]{fontenc}    
\usepackage{hyperref}       
\usepackage{url}            
\usepackage{booktabs}       
\usepackage{amsfonts}       
\usepackage{nicefrac}       
\usepackage{microtype}      
\usepackage{xcolor}         
\usepackage[pdftex]{graphicx} 
\usepackage[font=small,labelfont=bf]{caption}
\usepackage{physics}
\usepackage[english]{babel}

\usepackage{centernot}

\usepackage{algorithm}
\usepackage{algpseudocode}

\usepackage{biblatex} 
\newcommand{\indep}{\perp \!\!\! \perp}

\title{Causal Discovery for Gene Regulatory Network Prediction}

\author{%
  Jacob Rast \\
  \texttt{jrast@andrew.cmu.edu} \\
}

\begin{document}

\maketitle

\begin{abstract}
Biological systems and processes are networks of complex nonlinear regulatory interactions between nucleic acids, proteins, and metabolites. A natural way in which to represent these interaction networks is through the use of a graph. In this formulation, each node represents a nucleic acid, protein, or metabolite and edges represent intermolecular interactions (inhibition, regulation, promotion, coexpression, etc.). In this work, a novel algorithm for the discovery of latent graph structures given experimental data is presented.

\end{abstract}

\section{Introduction}
 The problem of representing a biological process of interest in a graphical structure is under active investigation and stands as one of the grand challenges in biology. 

Since RNA data is widely available, exclusive use of the transcriptome as a stand-in for other biomolecular expression levels is a well-studied formulation of the problem. Commonly, RNA microarray or single cell RNA-seq data is used to obtain gene expression levels of a large number of cells. This data is used to form a module of gene expression, expressed in a directed or undirected graph. If the “ground truth” or “gold standard” pathway is known, it can in turn be used to evaluate the network. 

While the human body contains hundreds of cell types and subtypes, each cell contains a nearly identical genome. It is the complex regulatory machinery that determines which proteins a cell expresses and at what time, giving rise to cellular diversity. Regulatory networks play an important role in determining whether or not a gene will be expressed. Canonical examples of the regulatory network include the lactose (or lac) operon \cite{jacob1961genetic} and Wnt pathway \cite{wnt}.

Discovery of gene regulatory networks is a longstanding biological problem \cite{wisdom_crowds}. The generation of high-throughput data relevant to the problem has allowed for the application of new approaches. A large influence in the field was a series of challenges issued from the National Center for Data to Health under the Dialogue on Reverse Engineering Assessment and Methods (DREAM) framework. DREAM challenges DREAM3 \cite{prill2010towards}, DREAM4 \cite{dream4} and DREAM5 \cite{wisdom_crowds} tasked participants with determining the network structures of a number of measurements of gene expression level from increasingly complex networks. These included both in simulated networks and those from well studied model organisms.

In the years since, a number of graphical methods have been developed expanding upon these approaches \cite{ni2018bayesian} \cite{pgm_grn}. One notable example is the use of a factor graph for the representation of gene regulatory networks, which was successful at recovering pathways found in E. coli from a well studied database. Another notable example was the use of reciprocal graphs, a highly general structure that is well suited to the study of gene expression given its ability to model loops, a drawback of traditional Bayesian networks. 

Prediction of the regulatory structure of biological pathways has wide-ranging applications such as the discovery of new cellular pathways, prediction of response to environmental changes, discovery of transcription factors for stem cell differentiation, and deeper understanding of cellular pathways.


\begin{figure}
  \centering
\includegraphics[scale = 0.25]{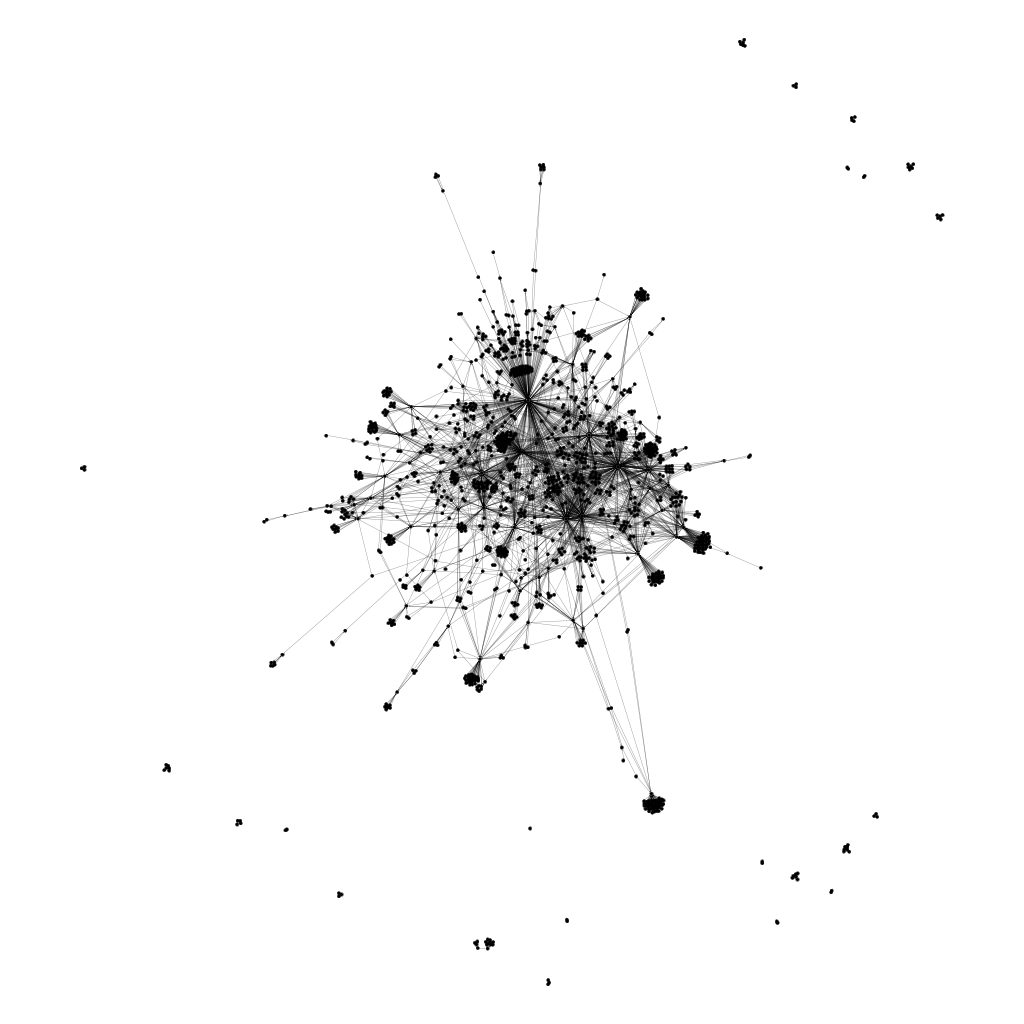}
\captionof{figure}{Full gene regulatory network for model organism E. coli}
\label{e_coli}
\end{figure}

\section{Background}

\subsection{Methods for modifying gene expression}
Several tools exist for manipulating the gene expression level in a target organism or cell. One of such techniques is the gene knockout, whereby the expression level of a target gene or set of genes is forced to 0. Typically this is achieved through genome editing whereby the DNA sequence or sequences coding for the set of genes to be knocked out is removed. This renders the organism incapable of expressing the RNA or protein encoded. While most commonly a single gene or dual gene knockout is performed, some studies have identified genome screens that allow for the study of combinations of up to three genes to be knocked out \cite{ZHOU2020108020}.  
A number of methods have been developed for gene engineering and knockout. Briefly, technologies such as CRISPR/Cas9 \cite{crispr} allow for the removal or insertion of a gene at any target location with minimal cost.
An alternative approach for manipulation of gene expression level in a target cell is the gene knockdown. In some instances it is advantageous to study the effect of perturbing a gene away from its steady state expression level rather than silencing its expression entirely. This can be achieved through the introduction of inhibitory molecules such as interfering RNA (RNAi), antisense DNA oligos, or inhibitory proteins.
Finally, some methods exists for inducing the expression of a target gene. While less robust and less well studied, some RNA species have been discovered that promote translation \cite{Carrieri2012-qa}.
In short, researchers have the tools to eliminate, increase, or decrease gene expression in a cell or organism. In this work algorithms were designed to study the experimental data resulting from each of these manipulations.

\subsection{Mathematical models of gene expression}

The use of tools to simulate gene expression data is invaluable to study the theory of reconstructing gene regulatory networks. A well-know tool for the simulation of gene expression is GeneNetWeaver \cite{Schaffter2011}. Briefly, GeneNetWeaver models a cell as a bipartite directed graph of proteins (transcription factors) and RNA. A series of ordinary differential equations or stochastic differential equations are used to relate the expression levels of the protein or RNA given a set of conditions. These conditions include biological factors such as RNA degradation rate, transcription factor-RNA affinity, transcription factor role (activation or deactivation of RNA transcription), etc. Interested readers may consult \cite{Kang2020} and \cite{Schaffter2011}. 

Steady-state solutions to the ODEs are used to generate baseline gene expression level. Experiments such as single gene knockout, multiple gene knockout, gene knockdown, gene knockup are simulated by intervening on the value of a set of variables and calculating the expression of the remaining variables under those conditions.

\subsection{Boolean Networks}
It is important to note the types of nonlinear dynamics that are frequently encountered in gene regulatory networks and that can be captured in ODE or SDE models. Frequently, proteins known as transcription factors will act in complexes in order to activate or inhibit the expression of a gene. The behavior whereby gene expression will only be affected by a complex and not at all by any subset of the complex is similar to the Boolean AND function. Simple Boolean network models of gene regulatory networks have been applied to capture this behavior \cite{Albert2003}. For more expressive power and the creation of synthetic datasets, success has been reported converting a Boolean network to a system of ordinary differential equations \cite{Pratapa2020}.


\section{Methods}

\subsection{Boolean Causal Discovery Algorithm}
In this section, the development, motivation, working theory, and proof of the Bool-PC algorithm are described.

\subsubsection{Test for independence}
Using observational data alone it is theoretically possible test for independence between variables using an approach such as mutual information. For small or simple networks without nonlinearities this approach can successfully reconstruct a directed graph \cite{Chow1968}. However, for larger or more complex networks this approach yields poor performance. Recognizing that the ability to manipulate the expression level of a gene is equivalent to  intervening on a variable (in the causal sense) a much more robust test for independence can be developed. 

We use the property that for two independent distributions, $P(A, B) = P(A)P(B)$ to write the following:
\begin{equation}
\label{eq:independence}
    A \indep B \implies P(A = a | B = b_1) = P(A = a | B = 0)
\end{equation}

In other words, if the probability distribution of A is unchanged by knocking out gene B, A and B are independent.

Additionally, note that for a directed network we have $A \indep B \centernot \implies B \indep A$ for independence as defined in this equation \ref{eq:independence}.

\subsubsection{Test for Conditional Independence}
This property is now extended to develop a test for conditional independence. A na\"ive extension would propose the following:

\begin{equation}
    A \indep B | C \implies P(A = a | B = b_1, C = c_1) = P(A = a | B = 0, C=c_1)
    \label{eq:cond_indep}
\end{equation}

Equation \ref{eq:cond_indep} holds for many probabilistic graphical models. It follows from the definition of conditional independence $P(A | B, C) = P(A | C)$. 

Importantly, equation \ref{eq:cond_indep} does not hold for the problem of gene regulatory network inference containing Boolean nonlinearitites. An illustrative counterexample can be found in the toy network depicted in figure \ref{fig:cond_indep}. From the graph structure, we can read the set of conditional independence statements $G27 \indep G26 | G25$ and $G27 \centernot \indep G24 | G25$.

In table \ref{table:1} we consider the effect of perturbing the value of G26 given G25 = 0. The value of G27 is unaffected, seemingly in accord with the conditional independence statements given in \ref{eq:cond_indep}. However, in table \ref{table:2} we consider the effect of perturbing G24 given G25 = 0. Under the same set of experimental conditions we observe a nearly identical result. The value of G27 is unaffected given G25=0, despite the graph structure giving conditional dependence. Finally, in \ref{table:3} the effect of perturbing G25 given G24 = 0 is given. Notice that again, no effect is observed on G27 from perturbations in G25 given G24 = 0, despite the graph structure describing conditional dependence. 

Given the Boolean AND inhibitory effect of G24 and G25 on G27, equation \ref{eq:cond_indep} does not capture the graph structure. This motivates the development of a test for conditional independence in Boolean causal graphs, given in \ref{cond_indep_bool}.

\subsection{Bool-PC}
Equation \ref{cond_indep_bool} is now used to develop a variant of the PC algorithm. Algorithm Bool-PC (\ref{alg:bool-pc}) is for use in Boolean-Nonlinear causal graphs suitable for use in gene regulatory network discovery.

Intuitively, the algorithm has the following logic. For any sets of variables X, Y, and Z if there is some flow of causality from X to Y, as shown in Step 1, we cannot prune an edge given Z blocks causal from X to Y if Y also blocks causal flow from Z to X, as these two conditions would imply no causal flow from Y, Z to X. 

\subsection{Bool-PC Proof of Correctness}
\begin{figure}
  \centering
  \includegraphics[scale = 0.6]{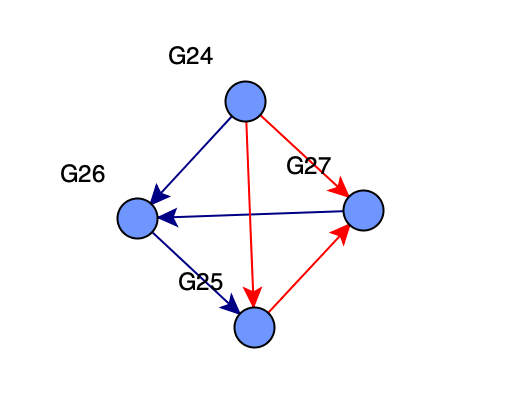}
  \caption{Toy model of conditional dependence and independence}
  \label{fig:cond_indep}
\end{figure}

The proof of correctness for Bool-PC follows from the reduction of SAT to 3SAT. For any 3 variables X, Y, Z we have shown exhaustively that Bool-PC can discover any graph structure, as shown in the toy example from Figure \ref{fig:cond_indep}. Given that the algorithm can reconstruct the Boolean relations between any 3 variables and that sets of 3 variables can be used to satisfy the Boolean relation of any arbitrary Boolean Satisfiability problem, we argue that Bool-PC is a correct algorithm for any arbitrary Boolean graph. From here the correctness of the PC algorithm applies.

 \begin{multline}
\label{cond_indep_bool}
     A \indep B | C \to P(A = a | B = b_1, C = c_1) = P(A = a | B = 0, C=c_1)\\
     \land P(A = a | B = b_1, C = c_1) = P(A = a | B = b_1, C=0) 
\end{multline}

\begin{algorithm}
\caption{Bool-PC Algorithm}\label{alg:bool-pc}
\begin{algorithmic}
\For{\texttt{all edges $E_{X, Y}$}} \Comment{Step 1}
\If{$P(X) = P(X | Y)$}
    \State \texttt{Remove edge $E_{X,Y}$}
\EndIf

\For{\texttt{Remaining edges $E_{X, Y}$ and sets of third variable Z}} \Comment{Step 2 to N}
\If{$P(X | Y, Z) = P(X | Y)$ and $P(X | Y, Z) = P(X | Z)$}
    \State \texttt{Remove edge $E_{X,Y}$}
\EndIf

\EndFor
\EndFor

\end{algorithmic}
\end{algorithm}

\begin{table}[h!]
\centering
\begin{tabular}{||c c c c c||} 
 \hline
 Conditions & G24 & G25 & G26 & G27 \\ [0.5ex] 
 \hline\hline
 Steady state & 0.541 & 0.464 & 0.110 & 0.112 \\ 
 G25 = 0 & 0.541 & 0.00 & 0.665 & 1.00 \\
 G25 = 0, G26=0 & 0.541 & 0.00 & 0.00 & 1.00 \\
 G25 = 0, G26 $\uparrow$ & 0.541 & 0.00 & 1.99 & 1.00 \\[1ex] 
 \hline
\end{tabular}
\caption{Experimental perturbations of G25 and G27}
\label{table:1}
\end{table}

\begin{table}[h!]
\centering
\begin{tabular}{||c c c c c||} 
 \hline
 Conditions & G24 & G25 & G26 & G27 \\ [0.5ex] 
 \hline\hline
 Steady state & 0.541 & 0.464 & 0.110 & 0.112 \\ 
 G25 = 0 & 0.54 & 0.00 & 0.67 & 1.00 \\
 G25 = 0, G24 = 0 & 0.00 & 0.00 & 0.09 & 1.00 \\
 G25 = 0, G24 $\uparrow$ & 2.07 & 0.09 & 0.69 & 1.00 \\[1ex] 
 \hline
\end{tabular}
\caption{Experimental perturbations of G25 and G24}
\label{table:2}
\end{table}

\begin{table}[h!]
\centering
\begin{tabular}{||c c c c c||} 
 \hline
 Conditions & G24 & G25 & G26 & G27 \\ [0.5ex] 
 \hline\hline
 Steady state & 0.541 & 0.464 & 0.110 & 0.112 \\ 
 G24 = 0 & 0.00 & 0.68 & 0.09 & 1.00 \\
 G24 = 0, G25 = 0 & 0.00 & 0.00 & 0.09 & 1.00 \\
 G24 = 0, G25 $\uparrow$ & 0.00 & 2.07 & 0.09 & 1.00 \\[1ex] 
 \hline
\end{tabular}
\caption{Experimental perturbations of G24 and G25}
\label{table:3}
\end{table}

\section{Results}
Performance of the GRNULAR + TopoDiffVAE algorithm, Bool-PC algorithm and comparison against the state of the art can be found in table \ref{table:4}.

\begin{table}[h!]
\centering
\begin{tabular}{||c c c ||} 
 \hline
 Metrics & AUPRC  & Training Time (seconds) \\ [0.5ex] 
 \hline\hline
 \text{Bool-PC (dual knockout)} & 0.741 & NA\\
 \text{Bool-PC (n-knockout, theoretical)} & 1.00 & NA\\
 \hline
\end{tabular}
\caption{Performance comparison of discovery algorithms on simulated GRN with 100 genes,
and 10 Transcription factors in a noise-free setting}
\label{table:4}
\end{table}



\section{Discussion and Analysis}
\subsection{Bool-PC}
First, it must be noted that the results present in table \ref{table:4} are for data generated without noise. The theoretically performance of Bool-PC on a network defined by a system of SDEs with noise (a more challenging problem that closer resembles the biological reality) cannot in general reach AUPRC of 1. Additionally, a major limitation of the Bool-PC algorithm is the experimental and time complexity. The PC algorithm has exponential time complexity, limiting its application to the discovery of full GRNs of thousands of nodes. Worse, the algorithm proposes an exponential number of gene knockout experiments which is both prohibitively expensive and technically infeasible. Bool-PC with dual gene knockout reliably captures well gene regulatory subnetworks of hundreds of nodes and very often perfectly captures networks of tens of nodes. This makes it a useful albiet limited tool. Use of the same conditional independence tests with more recent causal discovery algorithms such as the Grow-Shrink algorithm may expand the usefulness of the approach. Additionally, as gene sequencing costs shrink massive gene knockout panels will be increasingly feasible.




\section{Appendix}

\subsection{Access to Code}
All relevant data and code for the project can be found at the project's Github repository \href{https://github.com/jacobrast/grn_causal_discovery}{https://github.com/jacobrast/grn\_causal\_discovery}.

\printbibliography
\end{document}